\documentstyle[prl,aps,epsf]{revtex}
\advance\textheight by 0.14in
\advance\topmargin by -0.1in
\begin{document}
\draft

\twocolumn[\hsize\textwidth\columnwidth\hsize\csname@twocolumnfalse\endcsname

\title{Deconfinement in the Two Dimensional $XY$ Model} 
 
\author{H.A.\ Fertig}

\address{Department of Physics and Astronomy,
 University of Kentucky, Lexington, Kentucky 40506-0055}

\date{\today}

\maketitle

\begin{abstract} The unbinding of vortex-antivortex pairs for
the classical two-dimensional $XY$ model in a magnetic field
is studied.  A single such pair is connected by a string of
overturned spins, leading to linear confinement.  We show
that this system supports two phase transitions, one in which
closed strings proliferate, and a second in which vortices unbind.
The transitions are shown to be dual to one another, and are
remarkably continuous.  
Possible consequences
for a variety of systems are discussed.
\end{abstract}

]
{\it Introduction.} One of the most studied\cite{kadanoff} systems in condensed matter
physics is the two-dimensional $XY$ model,
\begin{equation}
H/k_B T= -K \sum_{<ij>} \cos(\theta_i-\theta_j) - h\sum_i \cos{\theta_i}.
\label{XYmodel}
\end{equation}
In Eq. \ref{XYmodel}, $\theta_i$ represents the angle with
respect to some reference direction of a planar spin located
at lattice site $i$, $\sum_{<ij>}$ denotes a sum over nearest 
neighbors, and $h$ represents the effect of a magnetic field
tending to align spins
along $\theta_i=0$.  This model receives so much
attention because it, and the closely related sine-Gordon
model\cite{gogolin}, describe a diverse range of 
two-dimensional classical, and one-dimensional quantum, systems.
The model supports
topological excitations in the form of vortices, which
for $h=0$ and high enough temperature
unbind via a Kosterlitz-Thouless (KT) transition.

Surprisingly little work has been done to study the
fate of this transition for $h>0$.
Early renormalization group (RG) studies\cite{jose} 
performed on the above model, with the last term replaced
by $h \sum_i \cos(p\theta_i)$ with $p$ an integer,
indicated a range of parameters
for which the term is irrelevant, presumably allowing
a KT transition to be realized.  For $p<4$ the symmetry-breaking
term is always relevant and the RG results are inconclusive.
Nevertheless, for $p=1$, the spins trivially exhibit
long-range order, and it is commonly supposed that
there can be no transition in this circumstance\cite{nelson_dg}.

In this article, we demonstrate that vortex unbinding does indeed
occur in the $XY$ model with a magnetic field, although the
nature of the transition is considerably changed from the
KT behavior.  
In the absence of fluctuations,
for large separations a state with a single
vortex-antivortex pair
will contain a string of reversed spins connecting them, leading
to a linear pair potential \cite{girv}.  
In the absence of fluctuations, the width of this string
is given by $\xi_0 = a_0 \sqrt{K/h}$, where $a_0$ is the underlying 
(square) lattice constant, and the energy per unit
length is $8\sqrt{Kh}~(k_B T/a_0)$ \cite{rajaraman}.  Even for small $K$,
pairs will be tightly bound and relatively isolated if
$h$ and $E_c$, the vortex 
core energy, are large.  As $h$ and $E_c$ decrease, 
vortices unbind in a two-step process: a first
phase transition occurs in which the vortices remain
bound in pairs, but the strings connecting them become
unbounded in length.  The resulting
state may be thought of as one in which closed strings
have proliferated.  This is
followed by a transition in which the strings
break open, with the endpoints being the vortices.  
This sequence has a dual description in terms
of a solid-on-solid model with screw dislocations \cite{chaikin}.
Interestingly, the vortex unbinding maps to a rough-flat
transition (equivalently, domain wall proliferation) in
the dual model, and string proliferation maps to dislocation unbinding.
The transitions are remarkable in that the free energy
is perfectly analytic as the system passes through them, in spite of
a diverging length scale 
and a clear change in the
low energy dynamics across the transition.
Fig. 1 illustrates
our proposed phase diagram for this system in the $E_c$ vs. $1/h$
plane, for a fixed value of $K \approx 1/2\pi$.

Such deconfinement transitions have consequences 
for a variety of systems, some of which we will briefly
describe towards the end of this article.

{\it String Model.}  We begin by explaining why in
a magnetic field, vortices come with strings attached.
For this purpose, we write the continuum version
of Eq. \ref{XYmodel},
\begin{equation} 
H =
\int d^2r \bigl\lbrace {1 \over 2}K|\vec{\nabla} \theta(\vec{r})|^2
-{h } \cos{\theta(\vec{r})} \bigr\rbrace,
\label{SGmodel}
\end{equation}
where we have chosen units in which $k_BT=1$ and $a_0=1$.
A vortex state is a local energy minimum of Eq. \ref{SGmodel},
subject to the constraint that
$\int_P d\vec{r} \cdot \vec{\nabla} \theta = 2\pi n$,
$n$ an integer,
along any closed path $P$ surrounding the vortex core.
Consider a state with a single vortex in a finite size system,
and a large path surrounding it.  
Because of the field, nearly all the spins 
along the path must
point in the preferred direction;
the $2\pi n$ rotation will arise as a localized kink.
If we choose the path parallel to $\vec{\nabla}\theta$
at the location of the kink and choose coordinates with the $\hat{y}$ direction
parallel to the path, $|\vec{\nabla} \theta(\vec{r})|^2
\approx (d\theta/dy)^2$, and we recognize the kink as
the soliton of the sine-Gordon model\cite{rajaraman}.
Since this kink must appear for any large path
enclosing the vortex, it is clear there is a string of
overturned spins running from the vortex to the system
boundary.   The existence of such string defects
in the context of the $XY$ model in a magnetic field
was noted in Ref.\cite{dieny}.

The unbinding transition for $h>0$ is driven by fluctuations
of the strings.  To see this,
we replace the partition function for Eq. \ref{XYmodel}
by the Villain model, in which the substitution
\begin{equation}
e^{C\cos\theta} \rightarrow \sum_{s=-\infty}^{\infty} e^{-C (\theta-2\pi s)^2/2}
\nonumber
\end{equation}
is made when cosines appear in the exponent.  
The Gaussian dependence of $\theta$ in the Villain model
allows some progress in evaluating the partition function
without sacrificing the critical properties of the $XY$ model,
since the two models share the same discrete periodic symmetry.
Using the techniques of Ref.\cite{jose}, the $\theta$ field
may be integrated out, and with some manipulation \cite{unpub}
the partition function for the Villain model may be written as 
$Z_{VM}=\sum_{\lbrace n({\bf r}),A({\bf r}) \rbrace} \exp[-H_{VM}]$,
with
\begin{equation}
H_{VM}=
\sum_{\bf r} \biggl[ {1 \over {2K}} \big|\vec{\nabla} n({\bf r}) + A({\bf r}) \hat{x} \big|^2
+ {1 \over {2h}} \biggl( {{\partial A} \over {\partial y}} \biggr)^2 \biggr] .
\label{VMmodel}
\end{equation}
In this expression $n({\bf r})$ and $A({\bf r})$ are integer degrees of freedom \cite{comA},
and derivatives should be understood as really meaning discrete differences;
e.g., $\partial A / \partial y \equiv A({\bf r}+\hat{y})-A({\bf r})$.

It is useful to think of the 
$n$ variables as residing at the centers of the square plaquettes formed
by the lattice sites, and the $A$ variables as residing on the nearest
neighbor bonds in the $\hat{y}$ direction.  
$H_{VM}$ may then be thought of as a solid-on-solid model
with screw dislocations \cite{chaikin}.  In this interpretation,
$n$ is an integer height variable, and non-vanishing derivatives
$\partial n / \partial x,~\partial n / \partial y$ locate domain walls
between different heights.   The important low energy excitations 
involving $A \ne 0$
contain line segments along the domain walls in which 
$\partial n / \partial x + A=0$, effectively removing part but not
all of a domain wall (see Fig. 2).  An endpoint of an ``A'' string --
where $\partial A / \partial y \ne 0$ --
may be identified as the core of a screw dislocation, and we see
the second term in $H_{VM}$ actually specifies its core energy
to be $1/2h$.  Our theory is thus effectively one in which the partition sum is
over graphs containing open strings, as might be expected from the
considerations discussed above.

Some comments are in order before we analyze $H_{VM}$. 
The model is closely related to one introduced in Ref. \cite{jose},
which expresses the partition sum in terms of 
two interacting vortex fields.  That model may be related
to $Z_{VM}$ by eliminating $n$ using the Poisson summation formula,
and representing the $A$ sum in terms of 
$\partial A / \partial y$ \cite{unpub}.  The resulting Hamiltonian is identical
to the vortex Hamiltonian of Refs. \cite{jose,nelson_dg} (see
Eq. 2.75b of Ref. \cite{nelson_dg}) with $E_c =0$.  A non-vanishing core
energy may be introduced in that representation, after which
the resulting theory enjoys a duality: upon interchanging $1/2h$ and $E_c$,
and changing $ K \rightarrow 1/4\pi^2 K $, the partition sum is identical to its original
form.   This observation will be important in obtaining the phase diagram.

{\it RG Analysis.} To make further progress we need 
a  model with continuous fields but the same symmetries as
$H_{VM}$.  A standard way to proceed \cite {cardy} is to exchange the integer
fields $n$ and $A$ for continuous variables $\phi$ and $a$, while
adding terms that favor integer values of the fields.  Taking the
continuum limit, our effective
Hamiltonian is
\begin{eqnarray}
H_{eff}=
&\int& d^2 r \biggl[ {1 \over {2K}} \big|\vec{\nabla} \phi({\bf r}) + a({\bf r}) \hat{x} \big|^2
+ {{\xi^2} \over {2K}} \biggl( {{\partial a} \over {\partial y}} \biggr)^2 \nonumber \\
-&y& cos\bigl( 2\pi \phi({\bf r}) \bigr) + y
-y_a cos\bigl( 2\pi a({\bf r}) \bigr) +y_a \biggr] . \\
\nonumber
\end{eqnarray}
For large $E_c$, $y=e^{-E_c}$ is the usual vortex fugacity \cite{com_resum}.
An important observation is that the low energy configurations
of $H_{eff}$ are highly analogous to those of $H_{VM}$,
even for small $y,~y_a$.  Domain walls in $n$ are analogous
to soliton configurations of $\phi$, and the $a$-field 
introduces local minima of $H_{eff}$ that
are analogous to open domain wall configurations.
An endpoint of an open wall
is analogous to a dislocation core in $H_{VM}$; by
matching the endpoint energy
to the dislocation core energy $1/2h$, one may assign
a value to $y_a$ such that there is a reasonable
mapping of the low energy configurations of $H_{VM}$ 
to those of $H_{eff}$.  A rough calculation yields
$y_a \approx 1/16h$ \cite{unpub}.

Our analysis proceeds from the small
$y,~y_a$ limit so that the cosine terms may be treated 
perturbatively.  We wish to integrate out short length
scale fluctuations in the partition function, which is
accomplished by integration of $\phi({\bf k}),a({\bf k})$
in the shell $\Lambda/b< |k_{x,y}| < \Lambda$, where
$b=e^l$ is the usual RG scale factor \cite{cardy} and 
$\Lambda = \pi$ is the momentum cutoff.
The scaling transformation takes the form 
${\bf k} = {\bf k^{\prime} } /b $, ${\bf r} = b {\bf r^{\prime} } $,
$\phi^{\prime}({\bf r^{\prime} }) = \phi({\bf r})$,
and $a({\bf r}) = a^{\prime}({\bf r^{\prime} }) /b$.
The renormalization of the $y_a \int d^2 r~cos\bigl[ 2\pi a({\bf r}) \bigr]$
term under this RG transformation
is unusual and leads to the 
remarkable critical properties of the deconfinement transition.
Since the argument of the cosine shrinks as the short wavelength
fluctuations are integrated out, one is led to expand the cosine
in a Taylor series,
\begin{equation}
y_a cos\bigl( 2\pi a \bigr) -y_a = \sum_{n=1}^{\infty} {y_{2n} \over {(2n)!}}
(-1)^n (2\pi a)^{2n},
\nonumber
\end{equation}
with the initial values $y_{2n}(l=0) = y_a$.  Upon rescaling,
power counting suggests $y_{2n} \sim b^{2-2n}$.
Thus we expect all the terms to
be irrelevant except the Gaussian one, which neither
grows nor shrinks.  The resulting fixed point Hamiltonian
has the form
\begin{equation}
H_{*}=
\int d^2 r \biggl[ {1 \over {2K}} \big|\vec{\nabla} \phi({\bf r}) + a({\bf r}) \hat{x} \big|^2
+ {1 \over 2} \rho   a({\bf r})^2
\biggr] .
\label{FP}
\end{equation}

The parameter $\rho$ plays an important role: it introduces
a finite energy per unit length for creating the ``$a$'' strings that
allow dislocation pairs to populate the system.  Thus $\rho$ is
effectively a (renormalized) string tension, and for $\rho>0$
dislocations are confined.  The behavior of $H_*$
reflects a qualitative difference when there is a finite 
tension: for $\rho=0$ there is a gapless mode for every 
value of $k_x$ when $k_y \rightarrow 0$, whereas for $\rho > 0$
there is only a single gapless mode at ${\bf k}=0$ \cite {com_rot}.
As we will see, the RG flows
determine which microscopic parameters allow a non-vanishing 
$\rho$.

In the vicinity of $H_{*}$, the scaling equations to lowest order
in $y,~y_n$ become
\begin{eqnarray}
{{dK} \over {dl}} & = & 0  \\
{{d\xi^2} \over {dl} } &=&-2 \xi^2  \label{rg_xi}\\
{{dy_{2n}} \over {dl} } &=& -(2n-2)y_{2n} -2\pi^2 {\cal L} (\rho,\xi) y_{2n+2} \label{y2n} \\
{{dy} \over {dl}} &=& \biggl(2-\eta \sqrt{K/\rho}\biggr) y . \label{yflow}\\
\nonumber
\end{eqnarray}
The last of these equations has been evaluated at $\xi(l)=0$, 
and $\eta$ is a number of order unity.  Note that $y_2 \equiv 4\pi^2 \rho$,
and the initial values $y_{2n}(l=0)$ are all given by $y_a$.
The function ${\cal L}$ arises from the momentum shell integral which
may be evaluated
exactly.  For small values of $\rho$, it has the form
$$
{\cal L} \approx {{K \Lambda^2} \over
{\pi \sqrt{K\rho(1+\xi^2\Lambda^2)}}}.
$$
A remarkable feature of Eqs. \ref{y2n} is that their solutions
are related by $y_{2n+2}(l) = y_{2n} e^{-2l} \equiv 
y_{2n} \xi(l)^2/\xi_0^2$ where $\xi_0=\sqrt{K/h}$ is the initial value
of $\xi(l)$.  Eqs. \ref{y2n} can then be rewritten in terms of a
single scaling equation,
\begin{equation}
{{d\rho} \over {d \xi^2}} = \pi^2 {\cal L}(\rho,\xi) \rho/\xi_0^2.
\label{rg_rho}
\end{equation}
Fig. 3 illustrates the trajectories resulting from Eq. \ref{rg_rho} 
for different initial values of $\xi_0$.  (The initial value of $y_a$ may be written
in terms of $\xi_0$
using the core energy matching discussed above.)   
For small values of $\xi_0$, 
$\rho(l)$ lies below the separatrix (shown as a heavy line), 
touches the $\rho=0$ axis
at a finite value of $\xi^2/\xi_0^2 \equiv e^{-2l^*}$, and remains on this axis as
it flows to the origin.  We associate $l_{scr}=a_0 e^{l^*}$ with a
screening length: for separations below this the dislocations appear
to be bound, while for larger separations the interaction is screened by
other dislocations, allowing for an unbinding transition.  This length
scale diverges as the phase transition is approached.
Above the separatrix, the flows end at a non-vanishing value of
$\rho$, which we identify as a renormalized string tension: the
dislocations are confined if $\xi_0$ is large enough.
Thus, the separatrix represents a deconfinement line, with dislocations
unbound if $\xi_0$ is smaller than some critical value $\xi_{cr}$.

Our RG analysis has demonstrated that for very large values of
$E_c$, there is a phase transition in which {\it dislocations} unbind.
The effect of {\it vortices} on this transition enters
at second order in $y$, and one can show \cite{unpub}
that this decreases the value of $\xi_0^2=K/h$ at which the transition
occurs as $E_c$ decreases from infinity.  On the
other hand, at $E_c/K=0,~K/h=0$, it may be shown that $H_{VM}$
has the character of a set of decoupled one-dimensional
solid-on-solid models \cite{unpub}, which should not have 
any phase transitions.  Thus, we presume the dislocation
unbinding line strikes the $K/h=0$ axis at a finite value value
of $E_c/K$, as shown in Fig. 1.

The dislocation confinement transition may be thought of as one in which
domain walls go from open strings to proliferated
closed loops.  The confined
phase is thus analogous to the rough phase of a solid-on-solid
model \cite{chaikin,log}.  For larger values of $y$ (smaller $E_c$),
a second transition to the smooth phase may occur \cite{kt}.  In the
dual of $H_{VM}$, this transition is equivalent to vortex
unbinding.  Because of the duality in the partition
sum, this transition has precisely the same character
as the dislocation unbinding transition, and we expect for
large values of $K/h$ that it should occur when 
$E_c/2\pi^2 K = \xi^2_{cr}$; i.e., at the same value one
finds for dislocation unbinding.  This is the origin of
the lower transition line in Fig. 1. It is interesting to note
that the confined phase is actually one in which
closed strings (loops) are proliferating.  In the dislocation language,
they are the domain walls associated with the
height $n$; in the dual representation (in which the
vortex degrees of freedom are explicit) they correspond
to strings of overturned spins.

A remarkable property of the deconfinement transition is that
it is perfectly continuous.
The ``accumulation point'' at $\rho=0,~\xi^2=0$ in Fig. 3 appears at the
end of a line of fixed points, but there is no relevant direction (in the RG sense)
leading away from it.  
Such relevant directions are the source of nonanalyticity in the free
energy in standard second order phase transitions \cite{cardy}.
Experimentally, this implies that thermodynamic measurements
cannot detect the deconfinement transition -- one must instead directly
probe the dislocations, vortices, or the correlation functions they
affect.  

Finally, we note that measurements of the vortex diffusion
constant in $XY$ model simulations 
have confirmed that vortex unbinding indeed occurs in this
system \cite{straley}.

{\it Applications.} The $XY$ system in a magnetic field
and several closely related models may be used to describe a large
number of systems, and the transition discussed in this
work has consequences for most of them.  
These include bilayer thin-film superconductors,
for which the {\it staggered} conductivity would be
controlled by a deconfinement transition;
bilayer crystals, for which dislocations
are analogous to vortices of the $XY$ model,
and deconfinement may lead to an
unlocking transition of the layers;
and double layer quantum Hall systems
at total filling factor $\nu=1$ \cite{girv},
where vortices carry physical charge $\pm e/2$
and deconfinement may lead to charge
fractionalization.  One dimensional quantum
systems (Luttinger liquids) are also often
described by models closely related to
Eqs. \ref{XYmodel} and \ref{SGmodel} \cite{gogolin},
and the transitions described in this work
have consequences under a variety of circumstances.
Finally, generalizations of the physics discussed
here to layered systems, including vortices
in high temperature superconductors, highly
anisotropic crystals, and quantum fluctuations
in 2+1 dimensional striped systems, are 
possible and are likely to lead to a variety
of new phenomena.

The author would like to thank G. Murthy,
J. Straley, D. Priour, A. Lau, and R. Golestanian for 
helpful suggestions and discussions.
The author is
indebted to Prof. Straley for providing results of simulations
at a crucial point in this research.
This work was supported by NSF Grant Nos. DMR-9870681
and DMR-0108451.

\begin{figure}
 \vbox to 5.0cm {\vss\hbox to 10cm
 {\hss\
   {\includegraphics{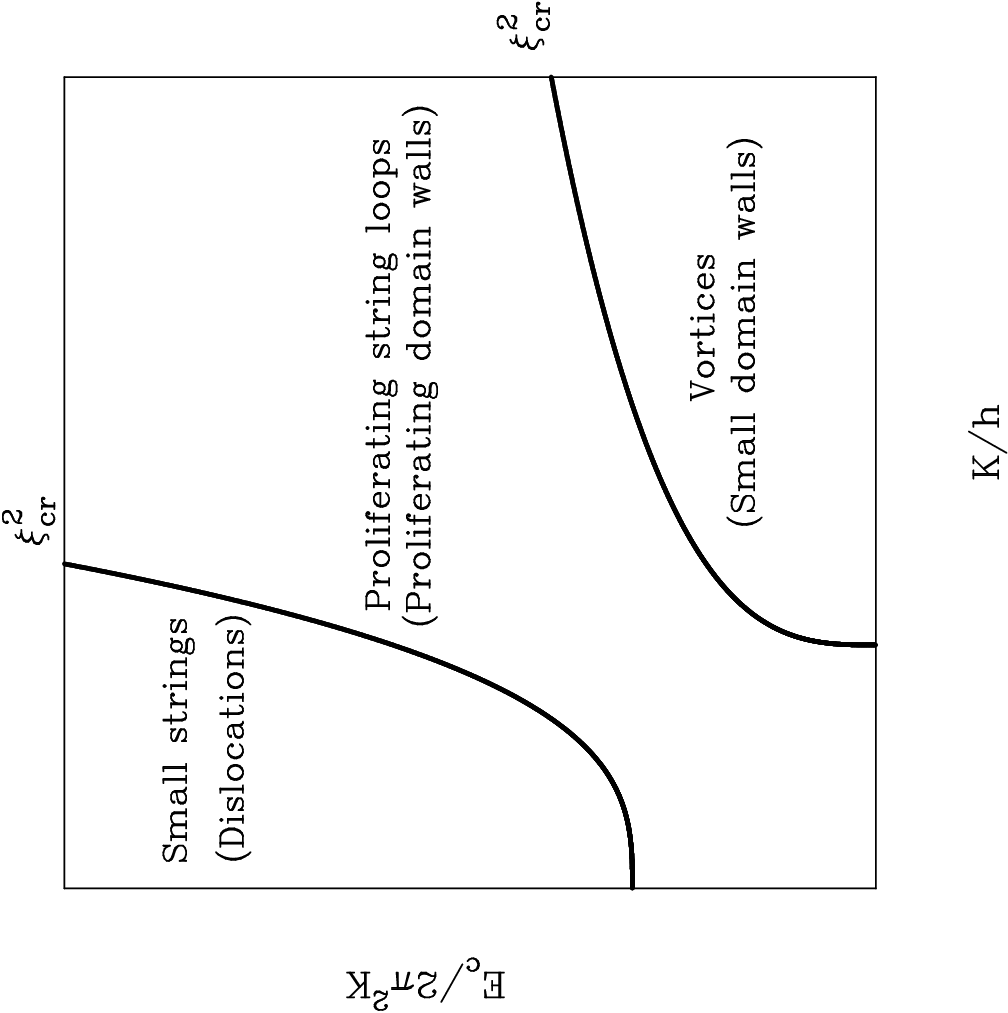}
   }
  \hss}
 }
\vspace{2mm}
\caption{Schematic phase diagram for $XY$ model in a magnetic field
for $K \approx 1/2\pi$.  Upper left corner
represents an ordered phase (unbound screw dislocations in dual representation).  
Middle phase contains proliferated
loops in both descriptions, and lower right corner contains unbound vortices 
(flat phase in dual representation.)}
\label{fig1}
\end{figure}

\begin{figure}
 \vbox to 5.0cm {\vss\hbox to 10cm
 {\hss\
   {\includegraphics{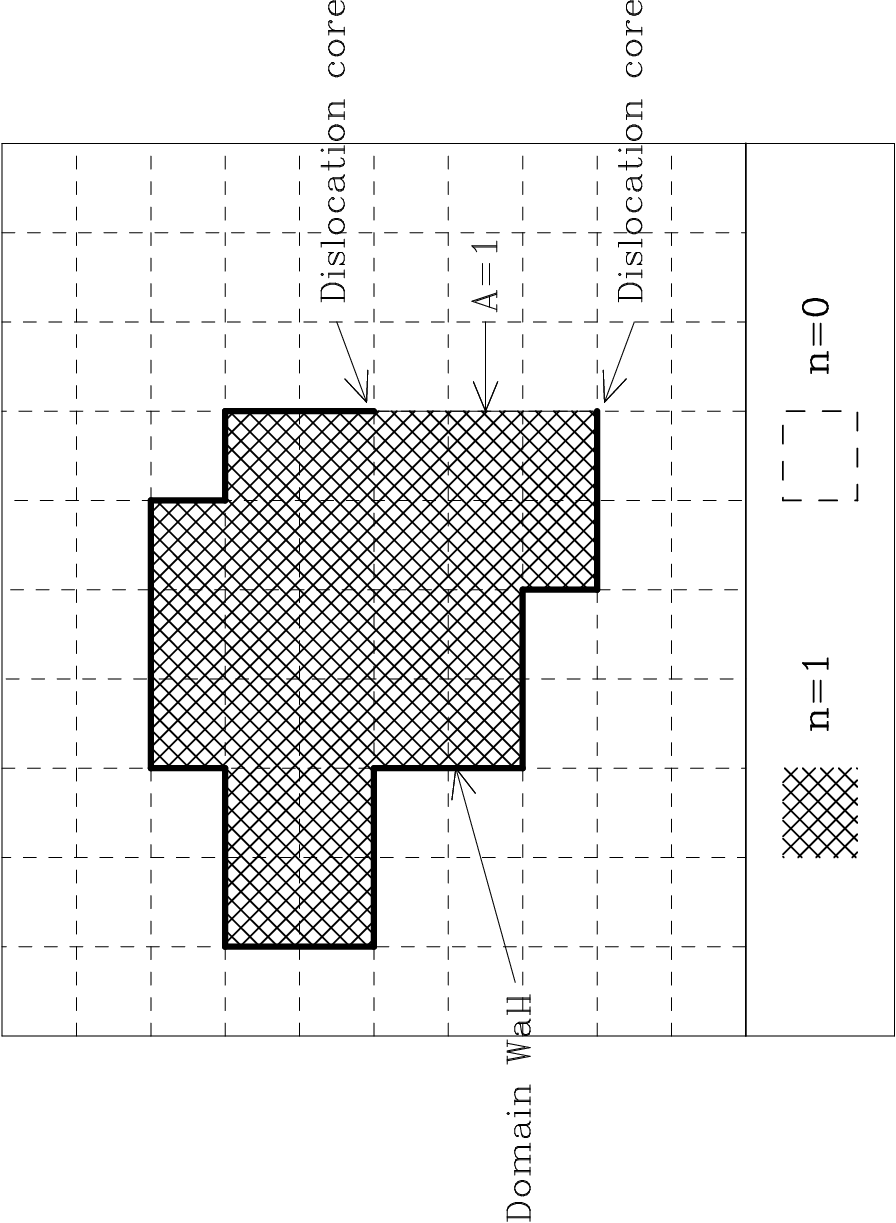}
   }
  \hss}
 }
\vspace{2mm}
\caption{A low energy configuration involving $A \ne 0$.  A region of $n=1$ 
(hatched squares) is
embedded in a surrounding $n=0$ region (white squares); heavy line represents a
domain wall.  The line segment with $A=1$ cancels the domain wall energy for
part of its length, leaving an open domain wall.}
\label{fig2}
\end{figure}

\begin{figure}
 \vbox to 5.0cm {\vss\hbox to 10cm
 {\hss\
   {\includegraphics{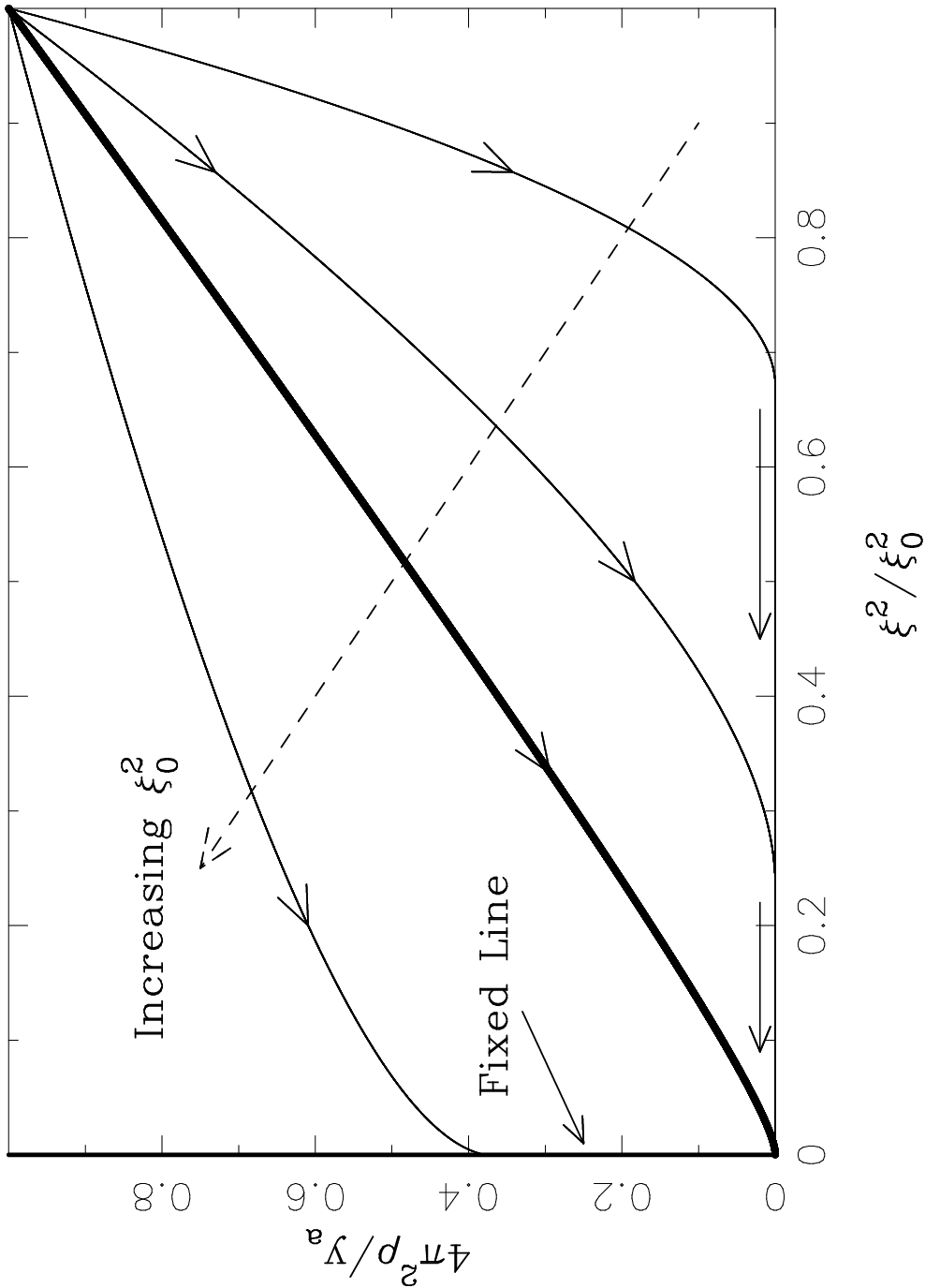}
   }
  \hss}
 }
\vspace{2mm}
\caption{RG flows for Eq. \ref{rg_rho}.  Left vertical axis is a fixed line, and heavy
line is a separatrix between flows that reach $\rho=0$ for finite $l$ and flow
to the origin as $l \rightarrow \infty$, and those that have $\rho>0$
at the end of their flow.  Finite values of $\rho$ in $H_*$ indicate confinement.}
\label{fig3}
\end{figure}

\begin{thebibliography}{10}

\bibitem{kadanoff}
See, for example, L.P. Kadanoff, {\it Statistical Physics: Dynamics,
Statics, and Renormalization}, (World Scientific, Singapore, 2000),
and references therein.

\bibitem{gogolin} A.O. Gogolin, A.A. Nersesyan, and A.M. Tsvelik,
{\it Bosonization and Stronly Correlated Systems,''} (Cambridge University
Press, Cambridge, 1998).

\bibitem{jose} J.V. Jos\'e, L.P. Kadanoff, S. Kirkpatrick, and
D.R. Nelson, Phys. Rev. B {\bf 16}, 1217 (1977).

\bibitem{nelson_dg} D.R. Nelson, ``Defect-Mediated Phase Transitions,''
in C. Domb and M.S. Green, {\it Phase Transitions and Critical Phenomena},
{\bf 7}, 1 (1983).

\bibitem{girv}
See article by S.M. Girvin and A.H. MacDonald,
in S. Das Sarma and A. Pinczuk Eds., {\it Perspectives in
Quantum Hall Effects}, (Wiley, New York, 1997).

\bibitem{rajaraman} R. Rajaraman, {\it Solitons and Instantons},
(North-Holland, New York, 1989).

\bibitem{chaikin} P.M. Chaikin and T.C. Lubensky, {\it Principles of
Condensed Matter Physics}, (Cambridge University Press, New York, 1995).

\bibitem{dieny} B. Dieny and B.A. Jones, J. Appl. Phys.
{\bf 69}, 5959 (1991).

\bibitem{unpub} H.A. Fertig, unpublished.

\bibitem{comA} There is a constraint in the partition sum that 
$A({\bf r})$ must vanish along some line
parallel to the $\hat{x}$ axis. This constraint does not need to be explicitly
enforced in the analysis that follows.


\bibitem{cardy} J. Cardy, {\it Scaling and Renormalization
in Statistical Physics,} (Cambridge University Press, New York, 1996).

\bibitem{com_resum} For small $y$, $y_a$, one may alternatively
derive $H_{eff}$ by applying the Poisson resummation formula
to the integer fields in $H_{VM}$ and retaining only the integers $0,\pm 1$.

\bibitem{com_rot} The apparent broken $\pi/2$ rotational symmetry
is an artifact of dislocation core energies depending on
whether the domain wall (steps in $n$) exits the core along the $x$ or
$y$ direction.  A less physical but qualitatively similar model can be formulated 
that respects the symmetry \cite{unpub}.

\bibitem{log} An interesting interpretation of this phase is that,
in the dual of $H_{VM}$, vortices are {\it logarithmically} confined \cite{unpub}.
To be consistent this means that the dislocations in $H_{VM}$
are also logarithmically confined in this phase.

\bibitem{kt} Eq. \ref{yflow} suggests that $H_{eff}$ supports a KT transition
for large enough values of $\rho$.   It is unlikely, however, that $H_{VM}$
accesses this transition, because the
the constraints of duality suggest a minimum value of $E_c$ is required
for vortex unbinding, in contrast to a $KT$ transition.  

\bibitem{straley} J.P. Straley and H.A. Fertig, unpublished.









\end{thebibliography}
\end{document}